# Simulation of induced radioactivity for Heavy Ion Medical Machine

XU Jun-Kui[1,2] SU You-Wu[1] LI Wu-Yuan[1] MAO Wang[1]
XIA Jia-Wen[1] CHEN Xi-Meng[2] YAN Wei-Wei[1] XU Chong[1]

*1 Institute of Modern Physics, Chinese Academy of Science, Lanzhou 730000, China*

*2 The School of Nuclear Science and Technology Lanzhou University, Lanzhou 73000, China*

**Abstract**: For radiation protection and environmental impact assessment purpose, the radioactivity induced by carbon ion of Heavy Ion Medical Machine (HIMM) was studied. Radionuclides in accelerator component, cooling water and air at target area which are induced from primary beam and secondary particles are simulated by FLUKA Monte Carlo code. It is found that radioactivity in cooling water and air is not very important at the required beam intensity and energy which is needed for treatment, radionuclides in accelerator component may cause some problem for maintenance work, suitable cooling time is needed after the machine are shut down.

## 1. Introduction

Nowadays radiation therapy is an important mean of treatment of tumor, more than 50% of all patients with localized malignant tumors are treated with radiation [1]. Many accelerators are built for medical purpose, in which heavy ion therapy is the most advanced treatment technology, and start at Bevalac facility at LBL in 1975 [2], and now there are 4 countries had launched the practice of heavy ion therapy. IMP (the Institute of Modern Physics, Chinese Academy of Sciences) had been started the research of biological effect of radiation with middle energy heavy ions since 1993, and start superficial tumor treatment of clinical research with 80MeV/u carbon ion beam in 2006. Two years later, deep treatment with higher energy heavy ion beam was beginning [3]. From 2010, HIMM was planned to establish in Lanzhou which is the special treatment device for carbon ion radiotherapy.

As a high-energy Heavy ion accelerator, induced radioactivity produced in accelerator and its beam-line components may cause exposure of maintenance workers, and makes the disposal of activated components difficult, thus may have certain radiation influence for environment. Meanwhile induced radioactivity in cooling water and air may affect not only the accelerator maintenance workers, but also the public and the environment through release. In this work, radionuclides induced by HIMM and their environmental impact are studied.

Research on induced radioactivity may trace back to very early since Curie and Joliot found the activation reaction in 1934[4], In the wake of improvement of accelerator, more kinds of particles, higher energy can be accelerated, research on induced radionuclides of accelerators continues in recent years[5-8]. Along with the heavy ion therapy put into practice, more peoples pay attention on environment radiation impact of heavy ion accelerator induced radioactivity, many reports can be found till very recently [9-10].

HIMM includes an ion source, low energy beam transport line, injector cyclotron, middle energy beam transport line, synchrotron (main accelerator), high energy beam transport line and 4 treatment rooms (Fig.1). Carbon ion was accelerated to 7MeV/u by the cyclotron and injected into synchrotron through middle energy beam transport line, and then the energy increased by the synchrotron range from 80MeV/u to 400MeV/u, delivered to treatment room through high energy beam line. The beam loss at extraction position and treatment room will result in induced radioactivity of the HIMM. In this paper induced radioactivity of accelerator component, cooling water and air around HIMM target area is studied with FLUKA MC Code. .

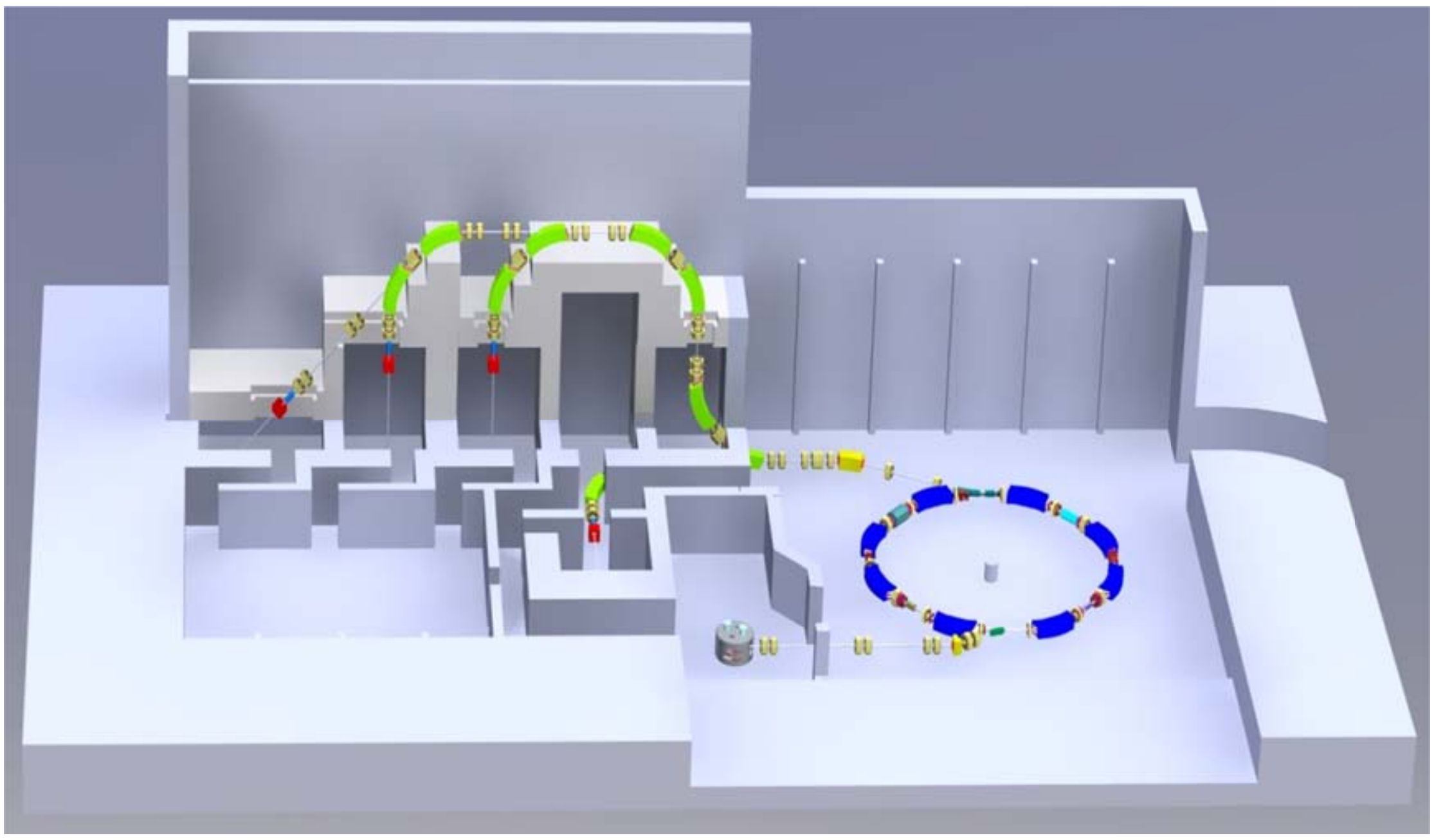

Fig.1 Schematic diagram of HIMM layout

## 2. Simulation

Heavy ion accelerator induced radioactivity includes two parts, those induced by primary ion beam is mainly exist in accelerator components and target, the activity is larger and concentrated on the beam loss position; radionuclides induced by secondary particles are widely distributed in shielding material, air and cooling water, their specific activity is very small. The radiation level of induced radioactivity depends on the accelerated ions, their energy, beam intensity, the material irradiated and the running time of accelerator, and so on.

In fact not many materials are used in the construction of accelerators. The most important are iron, stainless steel, copper, aluminum and aluminum alloys, and various hydrocarbons. Since heavier target materials produce more radionuclides, the copper was selected as the target material. FLUKA MC Code [11-12] is used in this work. The carbon ion energy is set as 400MeV/u, beam intensity is $1x10^8$pps, and the beam cross section diameter is 2mm. The copper target is a cylinder of 5cm in diameter and 10cm in thickness, and its axis coincides with the beam line, thus the thickness of target was enough to stop the primary beam. For calculation of radionuclides in cooling water, assuming the pipe axis is same as the copper target, its inner radio and outer radio was 20cm and 22cm respectively, and 38cm in length (Fig. 2). The size of Treatment Room is 4m×6m×6m, filled with dry air(Fig.3). The irradiated time is set to 30days. Compositions of the each material are shown in table 1.

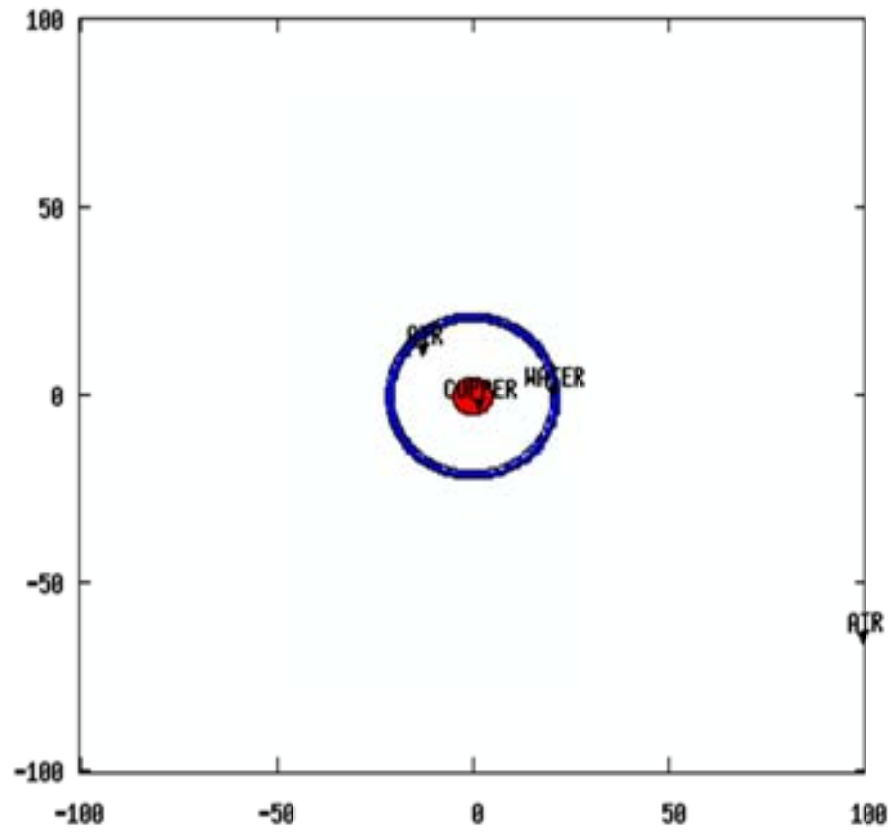

Fig.2 Schematic diagram of calculation model

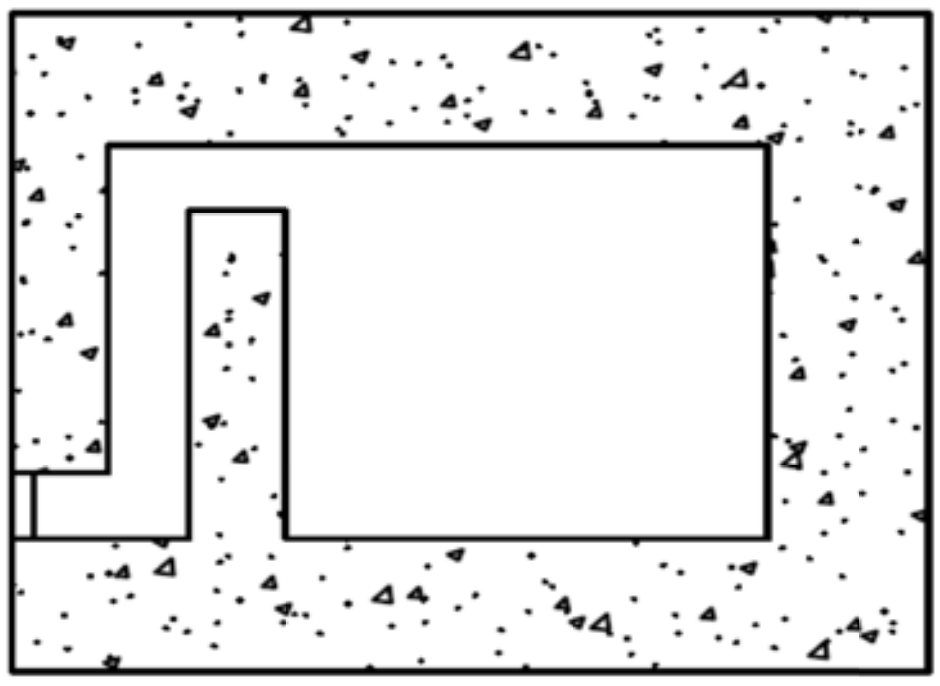

Fig.3 Schematic diagram of the treatment room

Table 1 Composition of each material

| Material | Element | Portion | Density (g/cm$^3$) |
|---|---|---|---|
| Copper | Cu | 1 | 8.9 |
| Water | H | 0.111111 | 1 |
| | O | 0.88889 | |
| Air | C | 0.000125 | 0.0012 |
| | N | 0.755267 | |
| | O | 0.231781 | |
| | Ar | 0.012827 | |

A series physical process such as spallation, neutron and γ capture, et al. nuclear reaction can be produced when Carbon ion beams bombardment on copper target. Physical process is descripted by the PHYSICS card which is supplied by FLUKA. For nuclear reactions induced by neutron activation, the LOW-NEUT low energy neutron transport must be activated. The IRRPROFI and DCYTIMES card is used to descript irradiation and cooling time of the target. Eventually, RESUNCLE card is chosen as the detector which can obtain radionuclides within the detector region.

**3. Results and discussion**

It can be seen from Fig.4 that there are more than two hundred induced radionuclides in cooper target after irradiation. But for the point of view of radiation protection, only a few radionuclides control the radiation field after irradiation. The important radionuclides for radiation protection are list in table 2. If there is more than one kind of radionuclide, only if the sum of the ratio of activity to its exempt value of each kind of the radionuclide is less than 1, it is exemptible [13]. The sum of the ratio of activity to their exempt value is greater than 3500, so the activity of the copper target is strongly. After irradiation, most radionuclides are short-life, though the total activity of the copper target is strong ($4.48\times10^9$Bq), one hour later activity declined to about half its original (Fig.5).

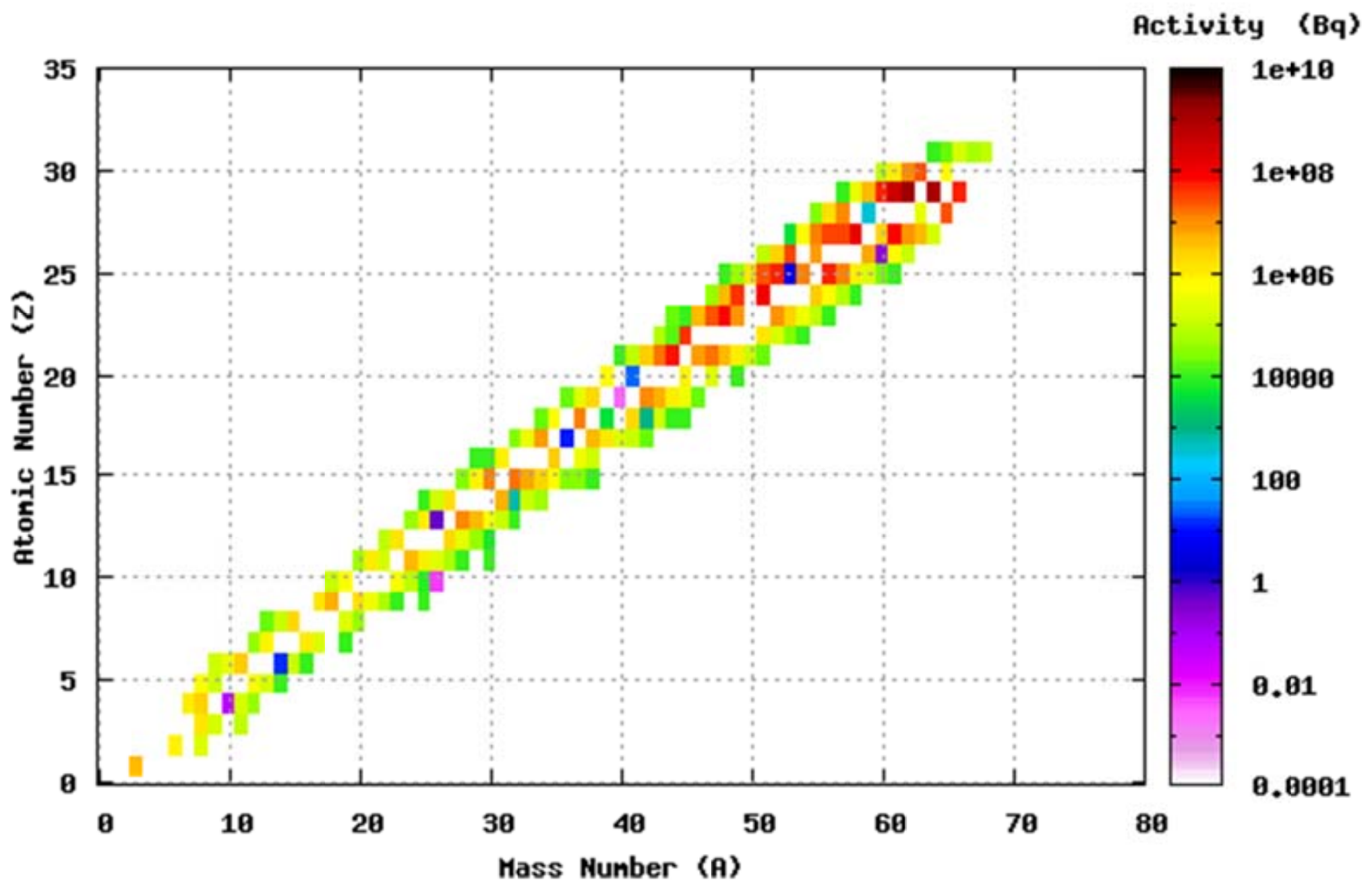

Fig. 4 The activity of radionuclides in copper target after irradiation

Table 2 Activity of main radionuclides in copper target

| radionuclide | Activity（Bq） | T1/2/ | Activity/exempt | radionuclide | Activity（Bq） | T1/2/ | Activity/exempt |
|---|---|---|---|---|---|---|---|
| $^{64}$Cu | $9.24\times10^{8}$ | 12.700 h | $9.24\times10^{2}$ | $^{46}$Sc | $7.59\times10^{6}$ | 83.79 d | $7.59\times10^{0}$ |
| $^{65}$Ni | $2.87\times10^{7}$ | 2.5172 h | $2.87\times10^{1}$ | $^{47}$Ca | $2.28\times10^{5}$ | 4.536 d | $2.28\times10^{-1}$ |
| $^{63}$Ni | $2.86\times10^{5}$ | 100.1 y | $2.86\times10^{3}$ | $^{45}$Ca | $9.23\times10^{5}$ | 162.61 d | $9.23\times10^{-2}$ |
| $^{65}$Zn | $5.92\times10^{5}$ | 244.26 d | $5.92\times10^{-1}$ | $^{42}$K | $1.21\times10^{7}$ | 12.36 h | $1.21\times10^{1}$ |
| $^{61}$Co | $7.52\times10^{7}$ | 1.650 h | $7.52\times10^{1}$ | $^{40}$K | $2.57\times10^{-3}$ | 1.277E+9 y | $2.57\times10^{-9}$ |
| $^{60}$Co | $2.17\times10^{6}$ | 5.2714 y | $2.17\times10^{1}$ | $^{41}$Ar | $1.84\times10^{6}$ | 109.34 m | $1.84\times10^{-3}$ |
| $^{57}$Co | $3.19\times10^{7}$ | 271.79 d | $3.19\times10^{1}$ | $^{37}$Ar | $1.38\times10^{7}$ | 35.04 d | $1.38\times10^{-1}$ |
| $^{56}$Co | $3.20\times10^{7}$ | 77.27 d | $3.20\times10^{2}$ | $^{38}$Cl | $4.32\times10^{6}$ | 37.24 m | $4.32\times10^{1}$ |
| $^{55}$Co | $1.50\times10^{7}$ | 17.53 h | $1.50\times10^{1}$ | $^{35}$S | $2.59\times10^{6}$ | 87.32 d | $2.59\times10^{-2}$ |
| $^{59}$Fe | $6.94\times10^{6}$ | 44.503 d | $6.94\times10^{0}$ | $^{33}$P | $6.12\times10^{6}$ | 25.34 d | $6.12\times10^{-2}$ |
| $^{55}$Fe | $7.51\times10^{6}$ | 2.73 y | $7.51\times10^{0}$ | $^{32}$P | $1.67\times10^{7}$ | 14.262 d | $1.67\times10^{2}$ |
| $^{52}$Fe | $2.32\times10^{6}$ | 8.275 h | $2.32\times10^{0}$ | $^{31}$Si | $5.00\times10^{6}$ | 157.3 m | $5.00\times10^{0}$ |
| $^{56}$Mn | $4.69\times10^{7}$ | 2.5785 h | $4.69\times10^{2}$ | $^{24}$Na | $3.92\times10^{6}$ | 14.959 h | $3.92\times10^{1}$ |
| $^{54}$Mn | $1.44\times10^{7}$ | 312.3 d | $1.44\times10^{1}$ | $^{22}$Na | $1.83\times10^{5}$ | 2.6019 y | $1.83\times10^{-1}$ |
| $^{52}$Mn | $5.08\times10^{7}$ | 5.591 d | $5.08\times10^{2}$ | $^{18}$F | $5.52\times10^{6}$ | 109.77 m | $5.52\times10^{0}$ |
| $^{51}$Mn | $2.47\times10^{7}$ | 46.2 m | $2.47\times10^{2}$ | $^{14}$C | $1.22\times10^{1}$ | 5730 y | $1.22\times10^{-6}$ |
| $^{51}$Cr | $1.03\times10^{8}$ | 27.7025 d | $1.03\times10^{1}$ | $^{15}$O | $2.08\times10^{6}$ | 122.24 s | $2.08\times10^{3}$ |
| $^{48}$V | $7.34\times10^{7}$ | 15.9735 d | $7.34\times10^{2}$ | $^{7}$Be | $1.05\times10^{6}$ | 53.12 d | $1.05\times10^{-1}$ |
| $^{47}$Sc | $1.69\times10^{7}$ | 3.3492 d | $1.69\times10^{1}$ | $^{3}$H | $4.38\times10^{6}$ | 12.33y | $4.38\times10^{-3}$ |
| $^{48}$Sc | $4.35\times10^{6}$ | 43.67 h | $4.35\times10^{1}$ | | | | |

Table 3 Activity of main air-borne radionuclides in treatment room

| radionuclide | Activity（Bq） | T1/2/ | Activity/exempt |
|---|---|---|---|
| $^{41}$Ar | $4.25\times10^{6}$ | 109.34 m | $4.25\times10^{-3}$ |
| $^{37}$Ar | $4.99\times10^{4}$ | 35.04 d | $4.99\times10^{-4}$ |
| $^{38}$Cl | $1.00\times10^{4}$ | 37.24 m | $1.00\times10^{-1}$ |
| $^{36}$Cl | $3.03\times10^{-2}$ | 3.0E+5 y | $3.03\times10^{-8}$ |
| $^{35}$S | $8.46\times10^{3}$ | 87.32 d | $8.46\times10^{-5}$ |
| $^{33}$p | $2.24\times10^{4}$ | 25.34 d | $2.24\times10^{-4}$ |
| $^{32}$p | $4.60\times10^{4}$ | 14.262 d | $4.60\times10^{-1}$ |
| $^{22}$Na | $2.16\times10^{2}$ | 2.6019 y | $2.17\times10^{-4}$ |
| $^{15}$O | $5.09\times10^{6}$ | 122.24 s | $5.09\times10^{-3}$ |
| $^{7}$Be | $1.59\times10^{6}$ | 53.12 d | $1.59\times10^{-1}$ |
| $^{3}$H | $8.88\times10^{4}$ | 12.33 y | $8.88\times10^{-5}$ |

Table 4 Activity of main radionuclides in cooling water

| radionuclide | Activity（Bq） | T1/2/ | Activity/exempt |
|---|---|---|---|
| $^{15}$O | $2.25\times10^{7}$ | 122.24 s | $2.25\times10^{-2}$ |
| $^{14}$C | $4.88\times10^{1}$ | 5730 y | $4.88\times10^{-6}$ |
| $^{7}$Be | $9.17\times10^{5}$ | 53.12 d | $7.17\times10^{-2}$ |
| $^{3}$H-3 | $4.77\times10^{4}$ | 12.33 y | $4.77\times10^{-5}$ |

From the above calculation results, the kind of radionuclides in the copper target is largest of the three materials. This is due to the process of the nuclear reaction. Radionuclides in copper are induced by carbon ions and secondary particles with copper. A idealization heavy ion interaction model known as Abrasion-Ablation model are detailed presentation by Gunzert-Marx K [14].

The calculated air-borne radionuclides are shown in table 4, $^{3}$H、$^{7}$Be、$^{15}$O、$^{41}$Ar etc. are the most important radionuclides which are mostly produced by the interaction of secondary neutron with air, $^{3}$H, $^{7}$Be is more interesting in radiation protection, other radionuclides are less important because of their shorter half-life or low yields. The harm of these radionuclides on human is beta, gamma irradiation and irradiation induced by inhalation. The total activity discharged to environment monthly in ventilation must be calculated. The assumption as following, tunnel ventilation frequency is 2 hours; the total activity within a dynamic ventilation cycle discharged to environment is equal the total activity after two hours irradiation ($A_2$). So the total activity discharge to environment one month is $A= A_2\times12\times30$ . The instantaneous activity of $^{3}$H, $^{7}$Be after irradiation 2 hours is 1.303Bq and 29.68Bq respectively, so one month is 469.08Bq and 10684.8Bq. This value is within the Limits prescribed of the State [15]. According to table 3 the total ratio of activity to their exempt value is less than 1 that is exempt; therefore its hazard to the environment can be neglected.

Radionuclides in water are from neutron induced $^{16}O$ spallation , the activation results are listed in table 5, $^{3}H$ and $^{7}Be$ is the most important radionuclides, their activity is $4.775\times10^{4}$Bq and $9.17\times10^{5}$Bq respectively. Cooling water is a closed-circuit circulation system, and assuming a daily loss is 0.5%, so $^{3}H$, $^{7}Be$ emissions in the environment for one month is very low，which is within the Limits prescribed of the State[15].

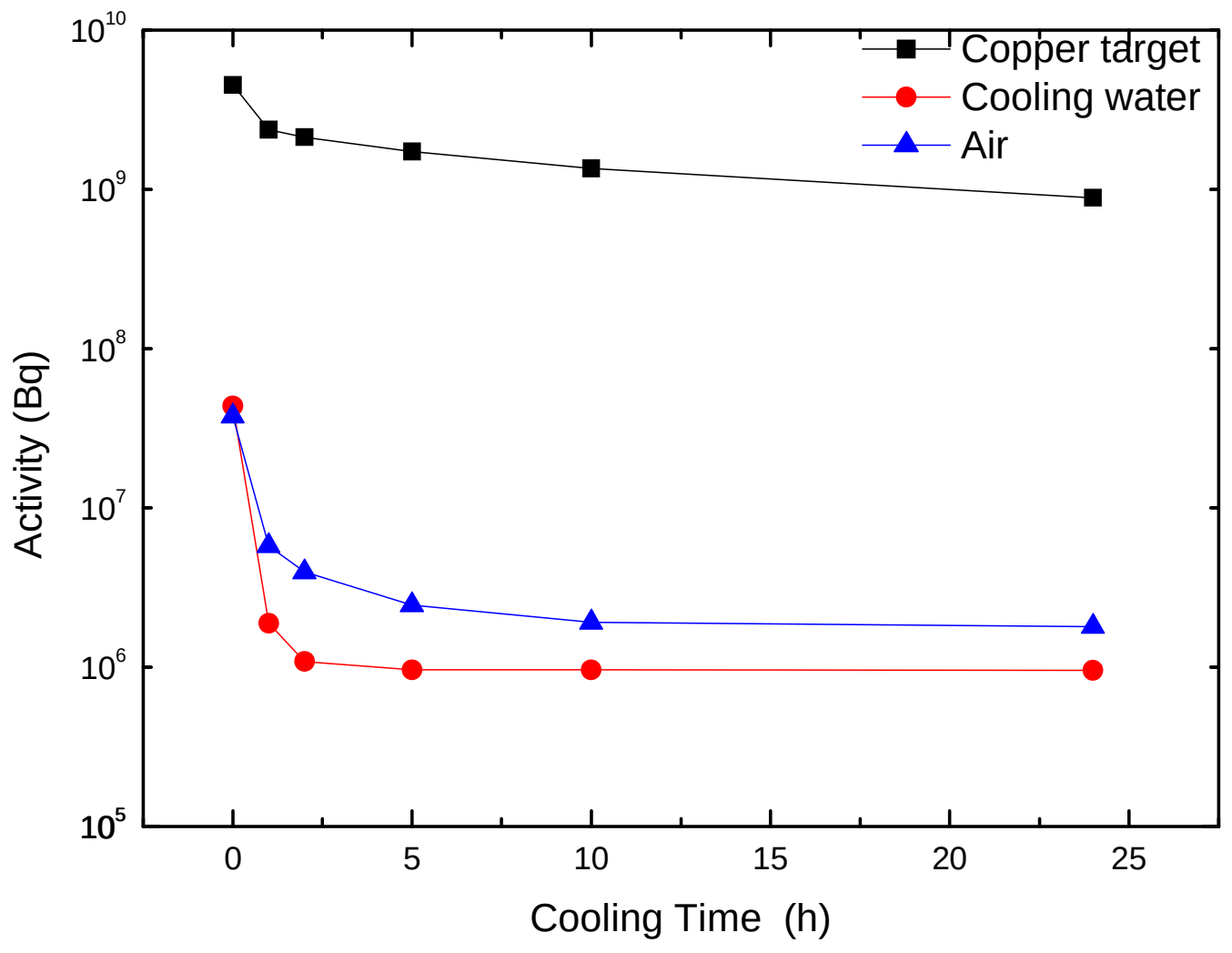

Fig.5 The decay of total activity of induced radioactivity

As the cooling time increasing, the total activity will decrease rapidly (Fig.5). The total activity in copper target will reduce to about half of its origin after one hour, and activity will reduce to one order of magnitude in air and cooling water. After cooling 5 hours, with the decay of short-life nuclides finished, the total activity decreases very slowly. Regardless of the type of activation material, radioactive decay law with the cooling time is similar. The initial point of the figure is the total activity of all radionuclides; the final straight portion is the only long-life radionuclides contribution. After the short-life radionuclides decay away quickly, and the total activity decay do not as fast as the original, thus forming a curved portion of the Fig.5. Therefore a suitable cooling time after accelerator shut down the maintenance worker can be entering the tunnel.

**4. Summery**

The radionuclides induced by HIMM were investigated with FLUKA MC Code in this paper. It is found that air and cooling water activation effect on the environment or the staff is not very important, the components of accelerator induced radioactivity has some influence on the maintenance personnel. After 1 hour the accelerator shut down, its activity can be reduced to the original 10% to 50%, therefore, after a period of cooling time the expose dose to maintenance worker will reduce rapidly. Some radiation protection measure may be need after a long time running since long live radionuclides was accumulated to a considerable amount. The data obtained in this work are fundamental radiation protection and environmental impact assessment.

**Reference:**

[1] Dieter Schardt, Thilo Elsässer. Rev. Mod. Phys. 2010, 82:383–425

[2] G T Y Chen, J R Castro, and, and J M Quivey, 1981, Annu Rev Biophys Bioeng. 1981,10:499-529

[3] XIAO Guo-Qing, ZHANG Hong, LI Qiang et al. Nuclear Physics Review, 2007, Vol.24, No.2 :86-87(in Chinese)

[4] Curie I, Joliot F. C.R. Acad. Sci., Paris, 1934, 198-254

[5] Fertman A, Mustafin E, Hinca R et al. Nucl. and Meth. , B260, 2007:579-591

[6] Strasik I, Mustanfin E, Fertman A et al. Nucl. and Meth., B266, 2008 :3443-3452

[7] Strasik I, Mustanfin E, Seidl Tet al. Nucl. and Meth., B268,2010: 573-580

[8] Wu Qing-Bao, Wang Qing-Bin, WU Jing-Min et al. CPC(HEP & NP), 2011, 35(6): 596–602

[9] Tujii H, Akagi T, Akahane K et al. Jpn., J. Med. Phys., 2009, 28(4):172-206

[10] Carbonez P, LaTorre F P, Michaud R et al. Nucl. and Meth. A 694, 2012:234–245

[11]Alfredo Ferrari, Paola R.Sala, Alberto Fasso et al. Fluka manual 2008

[12] Ballarini F, Battistoni G, Campanella M et al. Journal of Physics Conference Series 41, 2006:151-160

[13] GB-18871, Basic Standards of Safety of Radiation Sources. PRC (in Chinese)

[14] Gunzert-Marx K, Schardt D, Simon RS. Radiat. Prot. Dosim. 2004,110(1-4): 595-600 .

[15] GB 8703-88, Regulations for radiation protection. PRC (in Chinese)